\documentclass[prl,twocolumn,amsmath,amsfonts,amssymb,showpacs]{revtex4}
\usepackage{epsfig,psfrag,graphicx,bm}
\begin{document}

\title{Gamma Rays from Heavy Neutralino Dark Matter}
\author{Lars Bergstr\"om}
\email{lbe@physto.se}

\author{Torsten Bringmann}
\email{troms@physto.se}

\author{Martin Eriksson}
\email{mate@physto.se}

\author{Michael Gustafsson}
\email{michael@physto.se}

\affiliation{Department of Physics, Stockholm University, AlbaNova
  University Center, SE - 106 91 Stockholm, Sweden}

\date{August 8, 2005}

\pacs{95.35.+d, 11.30.Pb, 98.70.Rz}

\begin{abstract}
We consider the gamma-ray spectrum from neutralino dark matter
annihilations and show that internal bremsstrahlung of $W$ pair
final states gives a previously neglected source of photons at
energies near the mass of the neutralino. For masses larger than
about 1 TeV, and for present day detector resolutions, this results
in a characteristic signal that may dominate not only over
the continuous spectrum from $W$ fragmentation, but also over the
$\gamma\gamma$ and $\gamma Z$ line signals which are known to give
large rates for heavy neutralinos. Observational prospects thus seem
promising.
\end{abstract}

\maketitle

\newcommand{\pa}{\partial}
\newcommand{\vs}{\!\!\!\!\!}
\newcommand{\bl}{\Big[}
\newcommand{\br}{\Big]}
\newcommand{\nn}{\nonumber}
\newcommand{\de}{\delta}
\newcommand{\al}{\alpha}
\newcommand{\ga}{\gamma}
\newcommand{\ps}{\!\!\not\mbox{\hspace{-0.5mm}}p}
\newcommand{\ks}{\!\!\not\mbox{\hspace{-0.5mm}}k}
\newcommand{\qs}{\!\!\not\mbox{\hspace{-0.5mm}}q}
\newcommand{\be}{\begin{equation}}
\newcommand{\ee}{\end{equation}}
\newcommand{\neut}{\chi}
\hyphenation{micrOMEGAs higg-sino}

\section{Introduction}
One of the most favoured dark matter candidates
since 20 years or so, is the lightest supersymmetric particle
\cite{goldberg,ellisetal}. Most of the expected phenomenology has
been worked out, and is implemented in freely available, extensive
computer packages like {\sc DarkSusy} \cite{darksusy} and {\sc
micrOMEGAs} \cite{micromegas}. The vast majority of the analyses has
been performed in various constrained versions of the minimal
supersymmetric extension of the Standard Model (MSSM) (for reviews,
see e.g. \cite{neut}), where either radiative breaking of
supersymmetry or the GUT condition on gaugino masses (or both) have
been imposed. It is clear from all these studies that a suitable
neutralino candidate for dark matter is indeed available, with relic
density as measured by WMAP
($\Omega_\text{CDM} h^2 = 0.113 \pm 0.009$, where
$\Omega_\text{CDM}$ denotes the ratio of cold dark matter to
critical density and $h$ is the Hubble constant in units of
$100\text{~km} \text{\,s}^{-1} \text{\,Mpc}^{-1}$) \cite{wmap}, and
with great prospects of being detected either at the CERN LHC, or at
the various direct and indirect detection experiments of halo dark
matter that are currently operational or being constructed.

However, one may also ask the unpleasant but not unrealistic
question what happens if supersymmetry is not found at the LHC. If
supersymmetry is still present, that would probably mean that the
mass scale of the lightest supersymmetric particles is beyond the
kinematical reach of the accelerator. For neutralino masses at the
TeV scale, also the scattering cross section for direct detection
would necessarily be small, as would many of the rates (antiprotons,
positrons) for indirect detection. An exception seems to be
gamma-ray detection, firstly because rates do not fall off as
rapidly \cite{BUB} and secondly due to eminent new gamma-ray
telescopes, most clearly demonstrated by the recent spectacular
performance of HESS, in particular as regards the multi-TeV signal
that has been observed towards the galactic center \cite{hess}. With
such new astrophysical gamma-ray instruments of unprecedented size
and energy resolution either in operation \cite{hess,magic,cangaroo}
or under construction \cite{veritas,glast}, it is appropriate to
investigate possible levels of signals and spectral signatures of
heavy dark matter particle annihilation.

Most of the previous calculations have been carried out at
tree-level, with radiative corrections typically (and correctly)
believed to be at the few percent level. In some cases, radiative
corrections may on the contrary relieve the annihilation rate from
inhibiting factors having to do with the Majorana fermion property
of neutralinos and the fact that annihilation in the dark matter
halo effectively takes place at rest \cite{goldberg}. One example of
this is the radiative ``correction'' of $\neut\neut \to f\bar f$
through the emission of a photon in the final state. Here, the
first-order corrected cross section can be many orders of magnitude
larger than the tree-level result \cite{lbe89}.

A second example of an unexpectedly large cross section is that of
the second order, loop-induced $\gamma\gamma$ and $Z\gamma$
annihilation \cite{2gamma,zgamma,BUB}, where in the high mass, pure
higgsino (or wino \cite{ullio}) limit the branching ratio normalized
to the lowest order rate can reach percent level, despite the naive
expectation of being 2 to 3 orders of magnitude smaller. The origin
of this enhancement has only recently been fully understood, and is
explained by nonperturbative, binding energy effects in the special
situation of having very small (i.e.~galactic) velocities and very
large dark matter masses as well as small mass differences between
the neutralino and the lightest chargino \cite{hisano}.

In this Letter, we focus on gamma rays from neutralino annihilation
into charged gauge boson pairs and show that there is yet another,
previously neglected enhancement mechanism, appearing already at
first order in $\alpha_\mathrm{em}$: radiative processes containing
one photon in addition to the weak bosons in the final state, give a
new source of photons which peaks near the highest possible energy
(the mass of the neutralino). This turns out to be a very beneficial
effect for the potential detection of neutralinos with masses
$m_{\neut}\gtrsim1$~TeV, as the normally soft spectrum of continuous
photons coming from the fragmentation of $W$ or $Z$ bosons  (see,
e.g., \cite{BUB,hisano,fornengo})
gets a high-energy supplement with a clearly distinguishable
signature. This is reminiscent of the case of Kaluza-Klein dark
matter, where internal bremsstrahlung in annihilation processes with
charged  lepton final states dominates the gamma-ray spectrum at the
highest energies \cite{kk} (see also \cite{birkedal}).

\section{Gamma rays from neutralino annihilations}
In most models, the lightest stable supersymmetric particle is the
lightest neutralino, henceforth just ``the neutralino'', which is a
linear combination of the superpartners of the gauge and Higgs
fields, \be
  \neut\equiv\tilde\chi^0_1= N_{11}\tilde B+N_{12}\tilde W^3
+N_{13}\tilde H_1^0+N_{14}\tilde H_2^0\,. \ee In order not to
overclose the universe, a TeV-scale neutralino must generally have a
very large higgsino fraction,
$Z_h\equiv\left|N_{13}\right|^2+\left|N_{14}\right|^2$, if the usual
GUT condition $M_1\sim M_2/2$ is imposed; otherwise a heavy wino
would also be acceptable. In the following, we therefore focus on
higgsino-like neutralinos, with $Z_h\approx1$ and $N_{13}\approx\pm
N_{14}$ \footnote{ If the neutralino is a pure wino, one gets
identical results as for the (anti-) symmetric combination of
higgsinos that we consider here, apart from an overall factor of 16
that multiplies all quoted cross sections.
A pure bino state, on the other hand,  does not couple to $W$ at
lowest order at all.}. For the high masses we are interested in, the
annihilation rate into charged gauge bosons often dominates.
Internal bremsstrahlung in these final states are therefore of great
interest to investigate. Moreover, we note that all our results are
almost independent of the relative velocity $v$ of the annihilating
neutralino pair. Analytical expressions are therefore presented in
the limit of vanishing velocity, but should be applicable both at
the time of freeze-out ($v/c\sim1/6$) and to annihilating
neutralinos in the galactic halo today ($v/c\sim10^{-3}$).

\begin{figure}[ht]
  \includegraphics[angle=270,width=\columnwidth]{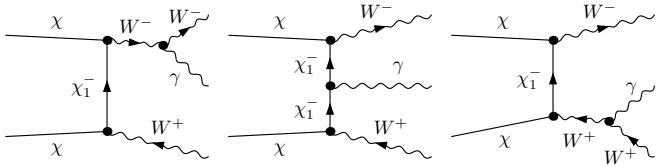}\\
  \caption{Contributions to $\neut\neut \rightarrow W^+ W^- \gamma$
  for a pure higgsino-like neutralino (crossing fermion lines are
  not shown).} \label{fig_feyn}
\end{figure}

For a pure higgsino, the only contribution to the lowest order
annihilation cross section into charged gauge bosons comes from a
$t$-channel exchange of a chargino; it is given by
\be
  (\sigma v)_{WW}=\frac{g^4}{32\pi}\frac{\left(m_{\neut}^2-m_W^2\right)\sqrt{1-m_W^2/m_{\neut}^2}}{\left(m_{\neut}^2+{m_{\chi_1^\pm}}^2-m_W^2\right)^2}\,,
\ee
where $m_{\chi_1^\pm}$ and $m_W$ are the lightest chargino and $W$ masses, respectively.

Let us now consider radiative corrections with a photon in the final
state in addition to the $W$ pair. Just as at lowest order, the
potential $s$-channel exchanges of $Z$ and Higgs bosons vanish, and
the only Feynman diagrams that contribute are shown in
Fig.~\ref{fig_feyn}. To zeroth order in $\epsilon \equiv
m_W/m_{\neut}$, and retaining a leading logarithmic term, the
resulting photon multiplicity is given by
 \vspace{0cm}
\begin{align}
\label{mult}
&\frac{\mathrm{d}N_{\gamma}^{W}} {\mathrm{d}x} \equiv \frac{\text{d}(\sigma v)_{WW\gamma}/\text{d}x}{(\sigma v)_{WW}} \nonumber \\
& \simeq \frac{\alpha_\mathrm{em}}{\pi} \bigg[ \frac{4(1-x+x^2)^2 \ln(2/\epsilon)} {(1-x)x} \nonumber \\
&-\frac{2(4-12x+19x^2-22x^3+20x^4-10x^5+2x^6)} {(2-x)^2(1-x)x} \nonumber \\
&+\frac{2(8-24x+42x^2-37x^3+16x^4-3x^5)\ln(1-x)} {(2-x)^3(1-x)x} \nonumber \\
&+\delta^2 \left( \frac{2x(2-(2-x)x)} {(2-x)^2(1-x)} +\frac{8(1-x)\ln(1-x)} {(2-x)^3} \right) \nonumber \\
&+\delta^4 \left( \frac{x(x-1)}{(2-x)^2}
+\frac{(x-1)(2-2x+x^2)\ln(1-x)} {(2-x)^3} \right)
\bigg]\,,
\end{align}
where $x\equiv E_\gamma / m_{\neut}$ and $\delta \equiv
(m_{\chi^\pm_1}-m_{\neut})/m_W$.

Several interesting features can be identified in this expression.
For large mass shifts $\delta$ the last two terms dominate. They
originate from longitudinally polarized charged gauge bosons in the
final state, which are forbidden in the lowest order process because
of the different CP properties of the initial and final state
\cite{dreesnojiri}. Remember that in the limit of vanishing relative
velocity, the initial state must be an $S$-wave with pseudoscalar
quantum numbers due to the Majorana nature of the neutralino. The
emission of a photon, on the other hand, will open up this channel
in the $^1S_0$ partial wave, potentially leading to very large cross
sections \footnote{ Unitarity is in general restored and therefore
forbids too large cross sections, which can be understood from the
equivalence theorem between longitudinal gauge bosons and would-be
Goldstone modes of the Higgs sector \cite{dreesnojiri}.}. However,
in supersymmetric scenarios with a heavy higgsino-like neutralino
one usually expects a mass shift $\delta < \epsilon$
\cite{dreesnojiri,hisano}, in which case the longitudinal part (last
two terms) can be neglected.

For small mass shifts, the cross section is instead dominated by the
production of transversely polarized gauge bosons. This results in a
peak in the spectrum at high energies that becomes more and more
pronounced for higher neutralino masses.  The appearance of this
peak can be understood by observing that for very heavy neutralino
masses the transversely polarized $W$ bosons can be treated as light
and thus behave in the same way as infrared photons radiated from
the neutralino/chargino line in Fig. \ref{fig_feyn}. The mechanism
that takes place is, in other words, an amusing reflection of QED
infrared behaviour also for $W$ bosons: The kinematical situation
when the photon and one of the $W$s leave the annihilation point
each with maximal energy, gets an enhancement, since it is
automatically accompanied by a very soft $W$. This is also reflected
in the symmetric appearance of the $x\rightarrow 0$ and $x
\rightarrow 1$ poles in the first terms of Eq.~(\ref{mult}).
\begin{figure}[thb]
  \includegraphics[width=\columnwidth]{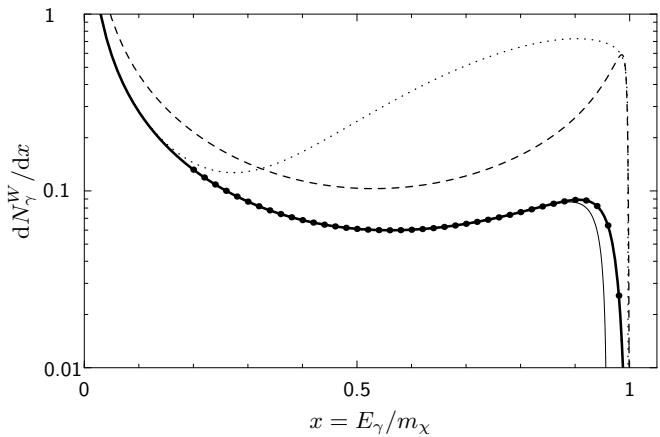}\\
  \caption{The photon multiplicity for the radiative processes
  $\neut\neut\rightarrow W^+W^-\gamma$. The dots represent the MSSM
  model of Table \ref{tab_susy}, as computed with the FormCalc
  package \cite{formcalc} for a relative neutralino velocity of
  $10^{-3}$.
  The thick solid line shows the full analytical result for the
  pure higgsino limit of the same model but with zero relative
  neutralino velocity. The thin solid line is the corresponding
  approximation as given in Eq.(\ref{mult}).
  Also shown, as dashed and dotted lines, are two pure higgsino
  models with a lightest neutralino (chargino) mass of 10 TeV
  (10 TeV) and 1.5 TeV (2.5 TeV), respectively. \label{fig_dN}}
\end{figure}

As an illustrative example, we have chosen a typical higgsino-like
MSSM model, fulfilling all experimental constraints, as specified in
Table \ref{tab_susy} (similar models are found in, e.g., the focus
point region of mSUGRA). The resulting photon spectrum from internal
bremsstrahlung of W pair final states is shown in Fig.~\ref{fig_dN}.
The symmetry around $x\sim0.5$ in the spectrum indicates the related
nature of the peak and the infrared divergence. For completeness, we
have also included a very high mass (10 TeV) higgsino model which
has received some attention recently \cite{hisano,Boudjema:2005hb}
(even though thermal production of such a neutralino in general
gives a too large $\Omega_\text{CDM}$, unless one allows for
finetuning of parameters like the psedudoscalar Higgs mass
\cite{Profumo:2005xd}). In addition, the case of a hypothetical
model with a very large mass shift is shown (where the contributions
from longitudinal $W$ bosons dominate at high energies).
\begin{table}[t]
\begin{ruledtabular}
   \begin{tabular}{cccccc|ccccc}
    $M_2$   &  $\mu$  &  $m_A$  &  $m_{\tilde f}$  &  $A_{f}$   &  $\tan \beta$  \;&
$m_\neut$  & $m_{\chi^\pm_1}$ & $Z_h$ & $W^\pm$  & $\Omega_\neut h^2$ \\
    \hline
    3.2 & 1.5 & 3.2 & 3.2 & 0.0 & 10.0 \;& 1.50 & 1.51 & 0.92 & 0.39 & 0.12\\
   \end{tabular}
\end{ruledtabular}
\caption{\label{tab_susy} MSSM parameters for the example model
shown in Fig.~\ref{fig_dN}-\ref{fig_smearedspectra} and the
resulting neutralino mass ($m_{\neut}$), chargino mass
($m_{\chi^\pm_1}$), higgsino fraction ($Z_h$), branching ratio into
$W$ pairs ($W^{\pm}$) and neutralino relic density ($\Omega_\neut
h^2$), as calculated with {\sc DarkSusy} \cite{darksusy} and {\sc
micrOMEGAs} \cite{micromegas}. Masses are given in units of TeV.}
\end{table}

Let us now consider those contributions to the gamma-ray spectrum
from the decay of heavy neutralinos that have been studied earlier.
Secondary gamma rays are produced in the fragmentation of the $W$
pairs, mainly through the decay of neutral pions. In addition to the
secondary spectrum, there are line signals from the direct
annihilation of a neutralino pair into $\ga\ga$ \cite{2gamma} and
$Z\ga$ \cite{zgamma}. Due to the high mass of the neutralino, these
lines cannot be resolved but effectively add to each other at an
energy equal to the neutralino mass.


\begin{figure}[t]
  \includegraphics[width=\columnwidth]{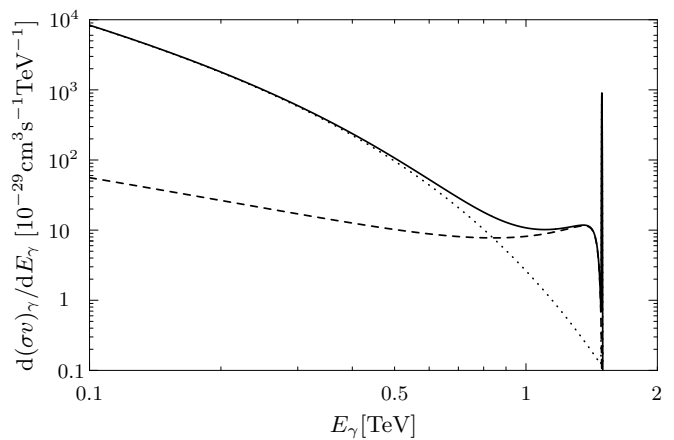}\\
  \caption{
The total differential photon distribution from $\neut\neut$
annihilations (solid line) for the MSSM model of Table
\ref{tab_susy}.
Also shown separately is the contribution from radiative processes
$\neut\neut\rightarrow W^+W^-\gamma$ (dashed), and the $W$
fragmentation together with the $\neut\neut\rightarrow\gamma\gamma,
\, Z\gamma$ lines (dotted).}
  \label{fig_allspectra}
\end{figure}

\begin{figure}[t]
  \includegraphics[width=\columnwidth]{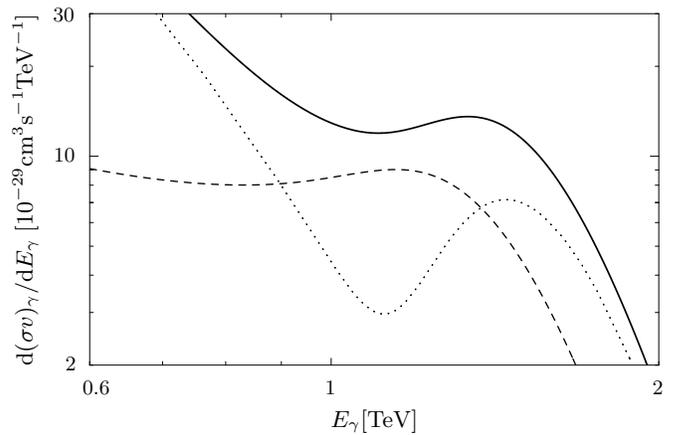}\\
  \caption{The same spectra as in Fig.~\ref{fig_allspectra}, as seen by a detector with an energy resolution of 15 percent.}
  \label{fig_smearedspectra}
\end{figure}

For comparison, again using the model of Table \ref{tab_susy},
Fig.~\ref{fig_allspectra} shows the contributions from secondary
photons \cite{hisano} and the line signals, as well as the new
source of photons from the internal bremsstrahlung diagrams of
Fig.~\ref{fig_feyn}.

The practical importance of the latter contribution can be
appreciated even more, when considering a finite detector resolution
of 15 \%, which is typical for atmospheric Cherenkov telescopes in
that energy range; the result is a smeared spectrum as shown in
Fig.~\ref{fig_smearedspectra}. One can see that, although the
strength of the $\ga\ga$ and $Z\ga$ lines already are surprisingly
large \cite{BUB}, the contribution from the internal bremsstrahlung
further enhances this peak by a factor of 2. The signal is also
dramatically increased at lower energies, thereby filling out the
``dip'' just below the peak; this latter effect will of course
become even more pronounced for better detector resolutions.

\section{Conclusions and Discussion}
In this Letter, we have presented important radiative corrections to
the gamma-ray spectrum from heavy neutralino annihilations. They
contribute a characteristic peak shape at the highest energies,
competing with the $\ga\ga$ and $Z\ga$ line signals in today's
detectors.

When it comes to predicting absolute gamma-ray fluxes, there is,
unfortunately, a great uncertainty which is primarily due to the
unknown dark matter clustering properties of the Milky Way halo.
Particle physics alone predicts, in lowest order perturbation
theory, a flux which falls roughly as $m_{\neut}^{-2}$. On the other
hand, the possible nonperturbative enhancement discussed in
\cite{hisano} is important for heavy neutralinos, and may give a
substantial boost to the gamma-ray signal. The process discussed
here should benefit from a similar boost as the line signal treated
in \cite{hisano}, since the enhancement is due to binding effects in
the initial state. Thus, even if the radiated $W$ boson is soft
compared to the initial state total mass, it still causes a
virtuality of one of the neutralinos which is much greater than the
binding energy. Therefore, the factorization of the process into
long distance and short distance kernels performed in \cite{hisano}
should be valid also here, and our curves for the ratios of cross
sections ($\mathrm{d}N_{\gamma}^{W} / \mathrm{d}x$) should not be
affected much.

Since the absolute gamma-ray flux - although not unlikely
considerable in size - is hard to predict, it is rather the spectral
shape that eventually may separate a dark matter signal from the
background. The HESS observations of the galactic center, e.g., show
a power-law energy spectrum with a spectral index of about $-2$ up
to at least 10 TeV \cite{hess}, and for Higgsinos as heavy as that,
we expect a spectrum that is too hard to explain the full data set
\cite{Profumo:2005xd}. Nevertheless, even a lighter higgsino could
still partly contribute; once one has access to better statistics, a
characteristic distortion in the spectrum would then be
distinguishable at the dark matter particle's mass. One should
furthermore bear in mind that the best prospects for detection might
therefore not be found near the galactic center, but rather for
sources with a low or at least well understood background. Examples
for this could be nearby dark matter clumps or intermediate mass
black holes \cite{Bertone:2005xz}; a thorough analysis of the
detectional prospects for these candidates, however, is beyond the
scope of this Letter.

Finally, we note that the radiative corrections presented in this
Letter add to the total annihilation cross section, quite
independently of the relative velocity of the annihilating
neutralinos. This is relevant for any precise calculation of
neutralino relic densities and is therefore of relevance for
computer packages like {\sc DarkSusy} and {\sc micrOMEGAs}.

\smallskip
 We thank J.~Edsj\"o for useful discussions and support
with the {\sc DarkSusy} package. L.B.~is grateful to the Swedish
Science Research Council (VR) for support.


\end{document}